# Exploring the Use of Virtual Worlds as a Scientific Research Platform: The Meta-Institute for Computational Astrophysics (MICA)

S. G. Djorgovski[1,*], P. Hut[2,*], S. McMillan[3,*], E. Vesperini[3,*], R. Knop[*], W. Farr[4,*], and M. J. Graham[1,*]

[1] California Institute of Technology, Pasadena, CA 91125, USA
[2] The Institute for Advanced Study, Princeton, NJ 08540, USA
[3] Drexel University, Philadelphia, PA 19104, USA
[4] Massachusetts Institute of Technology, Cambridge, MA 02139, USA

{lead author email: george@astro.caltech.edu}

**Abstract.** We describe the Meta-Institute for Computational Astrophysics (MICA), the first professional scientific organization based exclusively in virtual worlds (VWs). The goals of MICA are to explore the utility of the emerging VR and VWs technologies for scientific and scholarly work in general, and to facilitate and accelerate their adoption by the scientific research community. MICA itself is an experiment in academic and scientific practices enabled by the immersive VR technologies. We describe the current and planned activities and research directions of MICA, and offer some thoughts as to what the future developments in this arena may be.

## 1 Introduction

Immersive virtual reality (VR), currently deployed in the form of on-line virtual worlds (VWs) is a rapidly developing set of technologies which may become the standard interface to the informational universe of the Web, and profoundly change the way humans interact with information constructs and with each other. Just as the Web and the browser technology have changed the world, and almost every aspect of

---

modern society, including scientific research, education, and scholarship in general, a synthesis of the VR and the Web promises to continue this evolutionary process which intertwines humans and the world of information and knowledge they create.

Yet, the scientific community at large seems to be at best poorly informed (if aware at all) of this technological emergence, let alone engaged in spearheading the developments of the new scientific, educational, and scholarly modalities enabled by these technologies, or even new ideas which may translate back into the better ways in which these technologies can be used for practical and commercial applications outside the world of academia. There has been a slowly growing interest and engagement of the academic community in the broad area of humanities and social sciences in this arena (see, e.g., [1, 2, 3, 4, 5], and references therein), but the "hard sciences" community has barely touched these important and potentially very powerful developments. While a few relatively isolated individuals are exploring the potential uses of VWs as a scholarly platform, the scientific/academic community as a whole has yet to react to these opportunities in a meaningful way. One reason for this negligence may be a lack of the real-life examples of the scientific utility of VWs. It is important to engage the scientific community in serious uses and developments of immersive VR technologies.

With this growing set of needs and opportunities in mind, following some initial explorations of the VWs as a scholarly interaction and communication platform [6, 7], we formed the Meta-Institute for Computational Astrophysics (MICA) [8] in the spring of 2008. Here we describe the current status and activities of MICA, and its long-term goals.

## 2 The Meta-Institute for Computational Astrophysics (MICA)

To the best of our knowledge, MICA is *the first professional scientific organization based entirely in VWs*. It is intended to serve as an experimental platform for science and scholarship in VWs, and it will be the organizing framework for the work proposed here. MICA is currently based in *Second Life* (SL) [9] (it initially used the VW of *Qwaq* [10]), but it will expand and migrate to other VWs and venues as appropriate. The charter goals of MICA are:

1. Exploration, development and promotion of VWs and VR technologies for professional research in astronomy and related fields.
2. To provide and develop novel social networking venues and mechanisms for scientific collaboration and communications, including professional meetings, effective telepresence, etc.
3. Use of VWs and VR technologies for education and public outreach.
4. To act as a forum for exchange of ideas and joint efforts with other scientific disciplines in promoting these goals for science and scholarship in general.

To this effect, MICA conducts weekly professional seminars, bi-weekly popular lectures, and many other regularly scheduled and occasional professional discussions and public outreach events, all of them in SL. Professional members of MICA include scientists (faculty, staff scientists, postdocs, and graduate students), technologists, and professional educators; about 40 people as of this writing (March 2009). A broader group of MICA affiliates includes members of the general public interested in learning about astronomy and science in general; it currently consists of about 100 people (also as of March 2009). The membership of both groups is growing steadily. We have been very proactive in engaging both academic community (in real life and in SL) and general public, in the interests of our stated goals. Both our membership and activities are global in scope, with participants from all over the world, although a majority resides in the U.S.

MICA is thus a testbed and a foothold for science and scholarship in VWs, and we hope to make it both a leadership institution and a center of excellence in this arena, as well as an effective portal to VWs for the scientific community at large. While our focus is in astrophysics and related fields, where our professional expertise is, we see MICA in broader terms, and plan to interact with scientists and educators in other disciplines as well. We also plan to develop partnerships with the relevant industry laboratories, and conduct joint efforts in providing innovation in this emerging and transformative technology.

The practical goals of MICA are two-fold. First, we wish to lead by example, and demonstrate the utility of VWs and immersive VR environments generally for scientific research in fields other than humanities and social sciences (where we believe the case is already strong). In that process, we hope to define the "best practices" and optimal use of VR tools in research and education, including scholarly communications. This is the kind of activity that we expect will engage a much broader segment of the academic community in exploration and use of VR technologies. Second, we hope to develop new research tools and techniques, and help lay the foundations of the informational environments for the next generation of VR-enabled Web. Specifically, we are working in the following directions:

**2.1 Improving Scientific Collaboration and Communication**

Our experience is that an immediate benefit of VWs is as an effective scientific communication and collaboration platform. This includes individual, group, or collaboration meetings, seminars, and even full-scale conferences. You can interact with your colleagues as if they were in the same room, and yet they may be half way around the world. This is a technology which will finally make telecommuting viable, as it provides a key element that was missing from the flat-Web paradigm: the human interaction. We finally have a "virtual water cooler", the collegial gathering work spaces to enhance and expand our cyber-workspaces.

VWs are thus a very *green technology*: you can save your time, your money, and your planet by not traveling if you don't have to. This works well enough already, at almost no cost, and it will get better as the interfaces improve, driven by the games and entertainment industry, if nothing else. This shift to virtual meetings can potentially save millions of dollars of research funding, which could be used for more productive purposes than travel to collaboration or committee meetings, or to conferences of any kind.

We have an active program of seminars, lectures, collaboration meetings, and free-form scholarly discussions within the auspices of MICA, and we are proactive in informing our real-life academic community about these possibilities. We offer coaching and mentoring for the novices, and share our experiences on how to best use immersive VR for scientific communication and collaboration with other researchers.

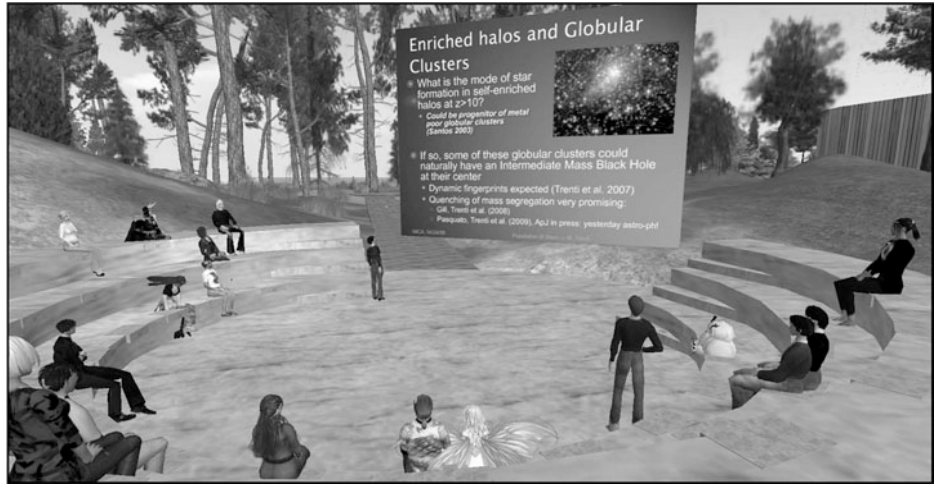

**Figure 1.** MICA members attending a regular weekly astrophysics seminar, in this case by Dr. M. Trenti, given in the *StellaNova* sim in SL. Participants in these meetings are distributed world-wide, but share a common virtual space in which they interact. [Preprint with the full-resolution color figures is available at http://www.mica-vw.org/wiki/index.php/Publications]

In addition, starting in a near future, we plan to organize a series of topical workshops on various aspects of computational science (both general, and specific to astrophysics), as well as broader-base annual conferences on science and scholarship in VWs, including researchers, technologists, and educators from other disciplines. These meetings will be either entirely based in VWs (SL to start), or be in "mixed reality", with both real-life and virtual environment gatherings simultaneously, connected by streaming media.

Genuine interdisciplinary cross-fertilization is a much-neglected path to scientific progress. Given that many of the most important challenges facing us (e.g., the global climate change, energy, sustainability, etc.) are fundamentally interdisciplinary in nature, and not reducible to any given scientific discipline (physics, biology, etc.), the lack of effective and pervasive mechanisms for establishment of inter-, multi-, or cross-disciplinary interactions is a serious problem which affects us all. One reason for the pervasive academic inertia in really engaging in true and effective interdisciplinary activities is the lack of *easy* communication venues, intellectual melting pots where such encounters can occur and flourish.

VWs as scientific interaction environments offer a great new opportunity to foster interdisciplinary meetings of the minds. They are easy, free, do not require travel, and the social barriers are very low and easily overcome (the ease and the speed of striking conversations and friendships is one of the more striking features of VWs). To this end, we will establish a series of broad-based scientific gatherings, from informal small group discussions, to full-size conferences. We note that once a VR environment is established, e.g., in a "sim" in SL, the cost (in both time and money) of organizing conferences is almost negligible, and the easy and instant worldwide access with no physical travel makes them easy to attend.

Thus, we have developed a dedicated "MICA island" (sim), named *StellaNova* [11] within SL. This is intended to be the Institute's home location in VWs; it is currently in SL as the most effective and convenient venue, but we will likely expand and migrate to other VW venues when that becomes viable and desirable. *StellaNova* is used as a staging area for most of our activities, including meetings, workshops, discussions, etc. It is intended to be a friendly and welcoming virtual environment for scholarly collaborations and discussions, very much in the tradition of *academe* of the golden age Athens.

A part of our exploration of VWs as scientific communication and collaboration platforms is an investigation in the mixed use of traditional Web (1.0, 2.0, ... 3.0?) and VR tools; we are interested in optimizing the uses of information technology for scientific communications generally, and not just exclusively in a VR context, although a VR component would always be present. We plan to evaluate the relative merits of these technologies for different aspects of professional scientific and scholarly interaction and networking – while the Web mechanisms may be better for some things, VWs may be better for others.

Finally, we intend to investigate the ways in which immersive VR can be used as a part of scientific publishing, either as an equivalent of the current practice of supplementing traditional papers with on-line material on the Web, or even as a *primary* publishing medium. Just as the Web offers new possibilities and modalities for scholarly publishing which do not simply mimic the age-old printed-paper media publishing, so we may find qualitatively novel uses of VWs as a publishing venue in their own right. After all, what is important is the content, and not the technical way in which the information is encoded; and some media are far more effective than others in conveying particular types of scholarly content.

### 2.2 A New Approach to Numerical Simulations

Immersive VR environments open some intriguing novel possibilities in the ways in which scientists can set up, perform, modify, and examine the output of numerical simulations. In MICA, we use as our primary science environment the gravitational N-body problem, since that is where our professional expertise is concentrated [12, 13, 14, 15, 16, 17], but we expect that most of the features we develop will find much broader applicability in the visualization of more general scientific or abstract data sets. Our goal is to create virtual, collaborative visualization tools for use by computational scientists working in an arbitrary VW environment, including SL [9], *OpenSim* [18], etc. Here we address interactive and immersive visualization in the numerical modeling and simulations context; we address the more general issues of data visualization below. For an initial report, see [40].

We started our development of in-world visualization tools by creating scripts to display a set of related gravitational N-body experiments. The gravitational N-body problem is easy to state and hard to solve: given the masses, positions, and velocities of a collection of N bodies moving under the influence of their mutual Newtonian gravitational interactions, according to the laws of Newtonian mechanics, determine the bodies' positions and velocities at any subsequent time. In most cases, the motion has no analytic solution, and must be computed numerically. Both the character of the motion and the applicable numerical techniques depend on the scale of the system.

Most of the essential features of the few-body problem can be grasped from studies of the motion of 3-5 body systems, in bound or scattering configurations. The physics and basic mathematics are elementary, and the required programming is straightforward. Yet, despite these modest foundations, such systems yield an extraordinarily rich spectrum of possible outcomes. The idea that simple deterministic systems can lead to complex, chaotic results is an important paradigm shift in many students' perception of physics. Few-body dynamics is also critically important in the determining the evolution and appearance of many star clusters, as well as the stability of observed multiple stellar systems.

These systems are small enough that the entire calculation could be done entirely within VWs, although we would wish to preserve the option of also importing data from external sources. This tests the basic capabilities of the visualization system – updating particles, possibly interpolating their motion, stopping, restarting, running backwards, resetting to arbitrary times, zooming in and out, etc.

The next level of simulation involves broadening the context of our calculations to study systems containing several tens of particles, which will allow us to see both the few-body dynamics and how they affect the parent system. Specifically, the study of binary interactions and heating, and the response of the larger cluster, will illustrate the fundamental dynamical processes driving the evolution of most star clusters. We will study the dynamics of systems containing binary systems, a possible spectrum of stellar masses, and real (if simplified) stellar properties. These simulations are likely to lie at the high end of calculations that can be done entirely within the native VW environments, and much of the data may have to be imported. The capacity to identify, zoom in on, and follow interesting events, and to change the displayed attributes of stars on the fly will be key to the visualization experience at this level.

The evolution of very large systems, such as galaxies, is governed mainly by large-scale gravitational forces rather than by small-scale individual interactions, so studies of galaxy interactions highlight different physics and entail quite different numerical algorithms from the previous examples. It will not be feasible to do these calculations within the current generation of VWs, or to stream in data fast enough to allow for animation, so the goal in this case will be to import, render, and display a series of static 3-D frames, which will nevertheless be "live" in the sense that particles of different sorts (stars, gas, dark matter, etc.) or with other user-defined properties can be identified and highlighted appropriately.

The choice of $N \sim 50,000$ is small compared to the number of stars in an actual galaxy, and it is more typical of a large star cluster. However, with suitable algorithms, galaxies can be adequately modeled by simulations on this scale, and this choice of N is typical of low-resolution calculations of galaxy dynamics, such as galaxy collisions and mergers, that are often used for pedagogical purposes. It also represents a compromise in the total amount of data that can be transferred into the virtual environment in a reasonable time. The intent here will be to allow users to visualize the often complex 3D geometries of these systems, and to explore some of their dynamical properties. This visualization effort in this case will depend on efficient two-way exchange of data between the in-world presentation and the external engine responsible for both the raw data and the computations underlying many aspects of the display.

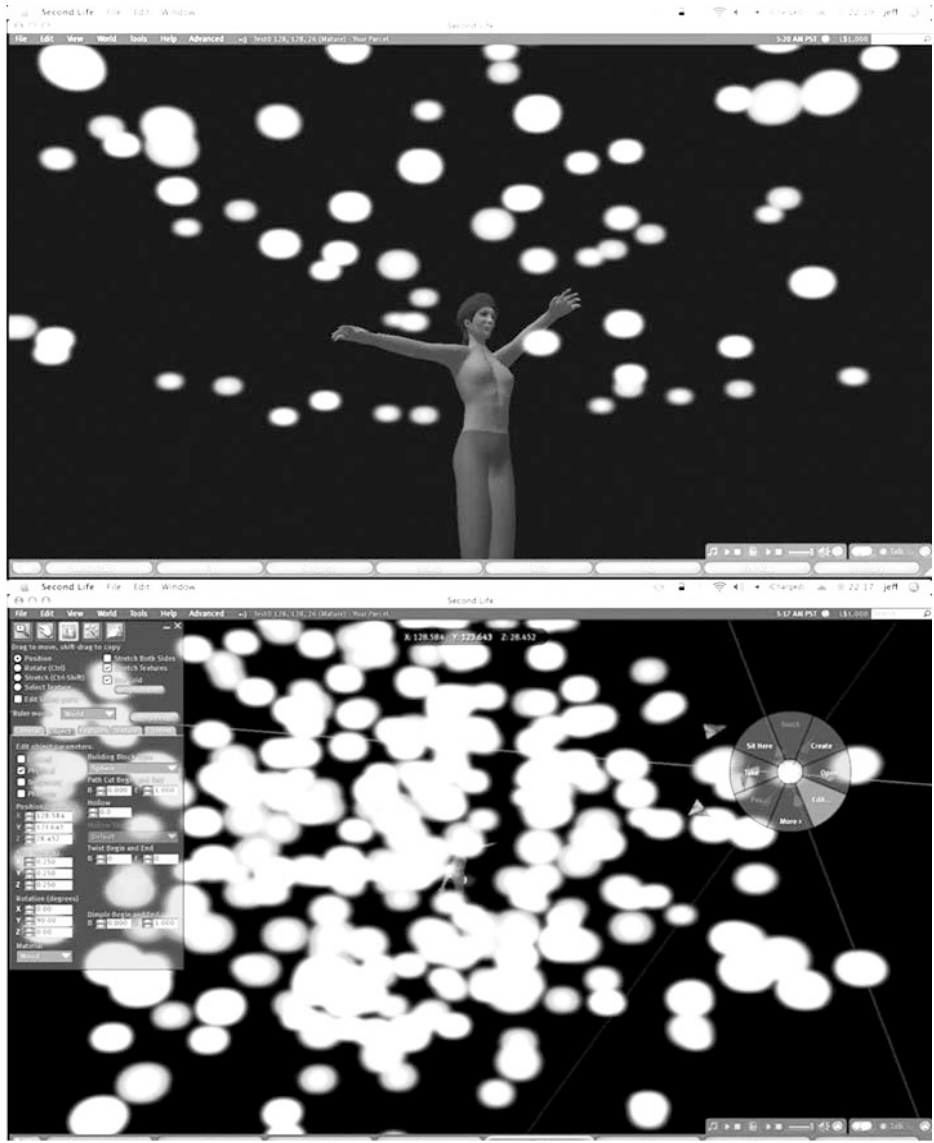

**Figure 2.** A MICA astrophysicist immersed in, and interacting with, a gravitational N-body simulation using the *OpenSim* environment. . [Preprint with the full-resolution color figures is available at http://www.mica-vw.org/wiki/index.php/Publications]

Our first goal is thus to explore the interactive visualization of simulations running within the VWs computational environments, thus offering better ways to understand the physics of the simulated processes – essentially the qualitative changes in the ways scientists would interact with their simulations. Our second goal is to explore the transition regime where the computation is actually done externally, on a powerful or specialized machine, but the results are imported into a VW environment, while the user feedback and control are exported back, and determine the practical guidelines as to how and when such a transition should be deployed in a real-life numerical study of astrophysical systems. The insights gained here would presumably be portable to other disciplines (e.g., biology, chemistry, other fields of physics, etc.) where numerical simulations are the only option of modeling of complex systems.

### 2.3 Immersive Multi-Dimensional Data Visualization

In a more general context, VWs offer intriguing new possibilities for scientific visualization or "visual analytics" [19, 20]. As the size, and especially the *complexity* of scientific data sets increase, effective visualization becomes a key need for data analysis: it is a bridge between the quantitative information contained in complex scientific measurements, and the human intuition which is necessary for a true understanding of the phenomena in question.

Most sciences are now drowning under the exponential growth of data sets, which are becoming increasingly more complex. For example, in astronomy we now get most of our data from large digital sky surveys, which may detect billions of sources and measure hundreds of attributes for each; and then we perform data fusion across different wavelengths, times, etc., increasing the data complexity even further. Likewise, numerical simulations also generate huge, multi-dimensional output, which must be interpreted and matched to equally large and complex sets of measurements. Examples include structure formation in the universe, modeling of supernova explosions, dense stellar systems, etc. This is an even larger problem in biological or environmental sciences, among others. We note that the same challenges apply to visualization of data from measurements, numerical simulations, or their combination.

How do we visualize structures (clusters, multivariate correlations, patterns, anomalies...) present in our data, if they are intrinsically hyper-dimensional? This is one of the key problems in data-driven science and discovery today. And it is not just the data, but also complex mathematical or organizational structures or networks, which can be inherently and essentially multi-dimensional, with complex topologies, etc. Effective visualization of such complex and highly-dimensional data and theory structures is a fundamental challenge for the data-driven science of the $21^{st}$ century, and these problems will grow ever sharper, as we move from Terascale to Petascale data sets of ever increasing complexity.

VWs provide an easy, portable venue for pseudo-3D visualization, with various techniques and tricks to encode more parameter space dimensions, with an added benefit of being able to interact with the data and with your collaborators. While there are special facilities like "caves" for 3D data immersion, they usually require a room, expensive equipment, special goggles, and only one person at a time can benefit from the 3D view. With an immersive VW on your laptop or a desktop, you can do it for free, and share the experience with as many of your collaborators as you can squeeze in the data space you are displaying, in a shared, interactive environment.

These are significant practical and conceptual advantages over the traditional graphics packages, and if VWs become the standard scientific interaction venue as we expect, then bringing the data to the scientists only makes sense. Immersing ourselves in our data may help us think differently about them, and about the patterns we see. With scientists immersed in their data sets, navigating around them, and interacting with both the data and each other, new approaches to data presentation and understanding may emerge.

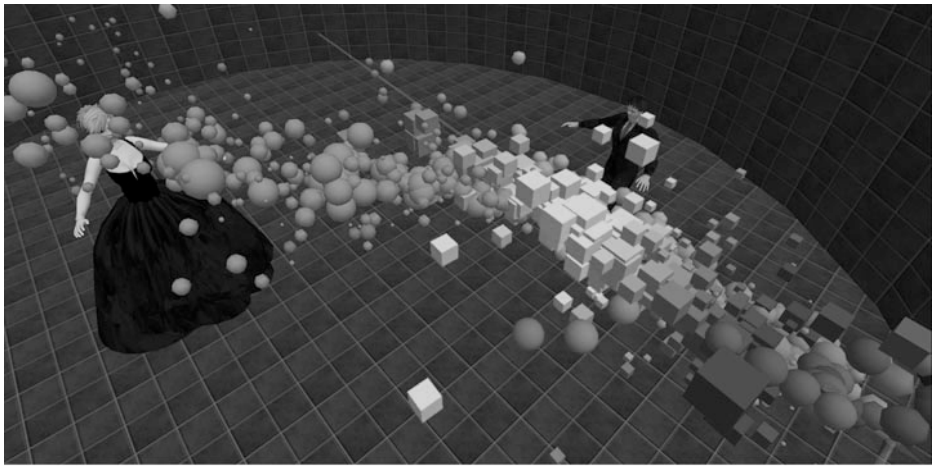

**Figure 3.** MICA scientists in an immersive data visualization experiment, developed by D. Enfield and S.G. Djorgovski. Data from a digital sky survey are represented in a 6-dimensional parameter space (XYZ coordinates, symbol sizes, shapes, and colors). [Preprint with the full-resolution color figures is available at http://www.mica-vw.org/wiki/index.php/Publications]

We have conducted some preliminary investigation of simple multi-dimensional data visualization scripting tools within SL. We find that we can encode data parameter spaces with up to a dozen dimensions in an interactive, immersive pseudo-3D display. At this point we run into the ability of the human mind to easily grasp the informational content thus encoded. A critical task is to experiment further in finding the specific encoding modalities that maximize our ability to perceive multiple data dimensions simultaneously, or selectively (e.g., by focusing on what may stand out as an anomalous pattern).

One technical challenge is the number of data objects that can be displayed in a particular VW environment; SL is especially limiting in this regard. Our next step is to experiment with visualizations in custom VW environments, e.g., using *OpenSim* [18], which can offer scalable solutions needed for the modern large data sets. However, even an environment like SL can be used for experimentation with modest-scale data sets (e.g., up to $\sim 10^4$ data objects), and used to develop the methods for an optimal encoding of highly-dimensional information from the viewpoint of human perception and understanding.

Additional questions requiring further research include studies of combined displays of data density fields, vector fields, and individual data point clouds, and the ways in which they can be used in the most effective way. This is a matter of optimizing human perception of visually displayed information, a problem we will tackle in a purely experimental fashion, using VWs as a platform.

The next level of complexity and sophistication comes with introduction of the time element, i.e., sequential visualization of changing data spaces (an obvious example is the output of numerical simulations of gravitational N-body systems, discussed in the previous section). We are all familiar with digital movies displaying such information in a 2-D format. What we are talking about here is *immersive 3-D data cinematography*, a novel concept, and probably a key to a true virtualization of scientific research. Learning how to explore dynamical data sets in this way may lead to some powerful new ways in which we extract knowledge and understanding from our data sets and simulations.

Implementing such data visualization environment poses a number of technical challenges. We plan experiment with interfacing of the existing visualization tools and packages with VW platforms: effectively, importing the pseudo-3D visualization signal into VWs, but with a goal of embedding the user avatar in the displayed space. We may be able to adopt some emergent solutions of this problem from the games or entertainment industry, should any come up. Alternatively, we may attempt to encode a modest-scale prototype system within the VW computational environments themselves. A hybrid approach may be also possible.

### 2.4 Exploring the OpenGrid and OpenSim Technologies

Most of the currently open VWs are based on proprietary software architectures, formats, or languages, and do not interoperate with each other; they are closed worlds, and thus probably dead ends. *OpenSimulator* (or *OpenSim*) [18] is a VW equivalent of the open source software movement. It is an open-source *C#* program which implements the SL VW server protocol, and which can be used to create a 3-D VW, and includes facilities for creating custom avatars, chatting with others in the VR environment, building 3-D content and creating complex 3-D applications in VW. It can also be extended via loadable modules or Web service interfaces to build more custom 3-D applications. *OpenSim* is released under a BSD license, making it both open source, and commercially friendly to embed in products.

To demonstrate the feasibility of this approach, we have conducted some preliminary experiments in the uses of *OpenSim* for astrophysical N-body simulations, using a plugin, *MICAsim* [21, 22]. We have modified the standard *OpenSim* physics engine as a plugin, to run gravitational N-body experiments in this VW environment. We found that it's practical to run about 30 bodies in a gravitational cold-collapse model with force softening to avoid hard binary interactions in the simulator, where a few simulator seconds corresponds to a crossing time. We believe that we could get another factor of two in N from code optimizations in this setting.

We will continue to explore actively the use of *OpenSim* for our work, and in particular in the arena or numerical simulations and visualization, and pay a close attention to the issues of avatar and inventory interoperability and portability. A start along these lines is *ScienceSim* [23]. Having an immersive VR environment on one's own machine can bypass many of the limitations of the commercial VW grids, such as SL, especially in the numbers of data points that can be rendered.

It is likely that the convergence of the Web and immersive VR would be in the form whereby one runs and manages their own VR environment in a way which is analogous to hosting and managing one's own website today. OpenSim and its successors, along with a suitable standardization for interoperability, may provide a practical way forward; see also [24].

### 2.5 Information Architectures for the Next Generation Web

One plausible vision of the future is that there will be a synthesis of the Web, with its all-encompassing informational content, and the immersive VR as an interface to it, since it is so well suited to the human sensory input mechanisms. One can think of immersive VR as the next generation browser technology, which will be as qualitatively different from the current, flat desktop and web page paradigm, as the current browsers were different from the older, terminal screen and file directory paradigm for information display and access. A question then naturally arises: what will be the newly enabled ways of interacting with the informational content of the Web, and how should we structure and architect the information so that it is optimally displayed and searched under the new paradigm?

To this effect, what we plan to do is to investigate the ways in which large scientific databases and connections between them (e.g., in federated data grid frameworks, such as the Virtual Observatory [25, 26, 27]) can be optimally rendered in an immersive VR environment. This is of course a universal challenge, common to all sciences and indeed any informational holdings on the Web, beyond academia.

Looking further ahead, many of the new scientific challenges and opportunities will be driven by the continuing exponential growth of data volumes, with the typical doubling times of ~ 1.5 years, driven by the Moore's law which characterizes the technology which produces the data [35, 36]. An even greater set of challenges is presented by the growth of data complexity, especially as we are heading into the Petascale regime [37, 38, 39]. However, these issues are not limited to science: the growth of the Web constantly overwhelms the power of our search technologies, and brute-force approaches seldom work.

Processing, storing, searching, and synthesizing data will require a scalable environment and approach, growing from the current "Cloud+Client" paradigm. Only by merging data and compute systems into a truly global or Web-scale environment – virtualizing the virtual – will sufficient computational and data storage capacity be available. A strong feature of such an environment will be high volume, frequent, low latency services built on message-oriented architectures as opposed to today's service-oriented architectures. There will be a heterogeneity of structured, semi-structured and unstructured data that will need to be persisted in an easily searchable manner. Atop of that, we will likely see a strong growth in semantic web technologies.

This changing landscape of data growth and intelligent data discovery poses a slew of new challenges: we will need some qualitatively new and different ways of visualizing data spaces, data structures, and search results (here by "data" we mean any kind of informational objects – numerical, textual, images, video, etc.). Immersive VR may become a critical technology to confront these issues.

Scientists will have to be increasingly immersed into their data and simulations, as well as the broader informational environment, i.e., the next generation Web, whatever its technological implementations are, simply for the sake of efficiency. However, the exponential growth of data volumes, diversity, and complexity already overwhelms the processing capacity of a single human mind, and it is inevitable that we will need some capable AI tools to aid us in exploring and understanding the data and the output of numerical models and simulations.

Much of the data discovery and data analysis may be managed by intelligent agents residing in the computing/data environment, that have been programmed with our beliefs, desires and intents. They will serve both as proxies for us reacting to results and new data according to programmed criteria expressed in declarative logic languages and also as our interface point into the computing/data environment for activities such as data visualization. Interacting with an agent will be a fully immersive experience combining elements of social networking with advances in virtual world software.

Thus, we see a possible diversification of the concept of avatars – as they blend with intelligent software agents, possibly leading to new modalities of human and AI representation in virtual environments. Humans create technology, and technology changes us and our culture in unexpected ways; immersive VR represents an excellent example of an enabling cognitive technology [28, 29].

## 2.6 Education and Public Outreach

VWs are becoming another empowering, world-flattening educational technology, very much like as the Web has already done. Anyone from anywhere could attend a lecture in SL, whether they are a student or simply a science enthusiast. What VWs provide, extending the Web, is the human presence and interaction, which is an essential component of an effective learning process. That is what makes VWs such a powerful platform for any and all educational activities which involve direct human interactions (e.g., lectures, discussions, tutoring, etc.). In that, they complement and surpass the traditional Web, which is essentially a medium to convey pre-recorded lectures, as text, video, slides, etc.

Beyond the direct mappings of traditional lecture formats, VWs can really enable novel collaborative learning and educational interactions. Since buildings, scenery, and props are cheap and easy to create, VWs are a great environment for situational training, exploration of scenarios, and such. Medical students can dissect virtual cadavers, and architects can play with innovative building designs, just moving the bits, without disturbing any atoms. Likewise, physicists can construct virtual replicas of an experimental apparatus, which students can examine, assemble, or take apart.

There is already a vibrant, active community of educators in SL [30, 31], and many excellent outreach efforts are concentrated in the SL SciLands virtual continent [32]. MICA's own efforts include a well-attended series of popular talks, "Dr.Knop talks astronomy" [33], which includes guest lecturers, as well as informal weekly "Ask an Astronomer" gatherings. We will continue with these efforts, and expand the range of our popular lectures.

Under the auspices of MICA, we are starting to experiment with regularly scheduled classes and/or class discussions in SL, and we will explore such activities in other VW environments as well. These may include an introductory astronomy class, or an advanced topic seminar aimed at graduate students. We will also try a hybrid format, where the students would read the lecture materials on their own, and use the class time for an open discussion and explanations of difficult concepts in a VW setting. We also plan to conduct a series of international "summer schools" on the topics of numerical stellar dynamics, computational science, and possibly others, in an immersive and interactive VW venue.

## 3  Concluding Comments

In MICA, we have started to build a new type of a scientific institution, dedicated to an exploration of immersive VR and VWs technologies for science, scholarship, and education, aimed primarily at academics in physical and other natural sciences. MICA itself is an experiment in the new ways of conducting scholarly work, as well as a testbed for new ideas and research modalities. It is also intended to be a gateway for other scholars, new to VWs, to start to explore the potential and the practical uses of these technologies in an easy, welcoming, and collegial environment.

MICA represents a multi-faceted effort aimed to develop new modalities of scientific research and communication using new technologies of immersive VR and VWs. We believe that they will enable and open qualitatively new ways in which scientists interact among themselves, with their data, and with their numerical simulations, and thus foster some genuine new "computational thinking" [34] approaches to science and scholarship.

We use the VWs as a platform to conduct rigorous research activities in the fields of computational astrophysics and data-intensive astronomy, seeking to determine the potential of these new technologies, as well as to develop a new set of best practices for scholarly and research activities enabled by them, and by a combination of the existing Web-based and the new VR technologies. In that process, we may facilitate new astrophysical discoveries. We also hope to generate new ideas and methods which will in turn stimulate development of new technological capabilities in immersive VR and VWs, both as research and communication tools, and in the true sense of human-centered computational engineering.

The central idea here is that immersive VR and VWs are potentially transformative technologies on par with the Web itself, which can and should be used for serious purposes, including science and scholarship; they are not just a form of games. By conveying this idea to professional scientists and scholars, and by leading by example, we hope to engage a much broader segment of the academic community in utilizing, and developing further these technologies.

This evolutionary process may have an impact well beyond the academia, as these technologies blend with the cyber-world of the Web, and change the ways we interact with each other and with the informational content of the next generation Web. While at a minimum we expect to develop a set of "best practices" for the use of VR and VWs technologies in science and scholarship, it is also possible that practical and commercial applications may result or may be inspired by this work. If indeed immersive VR becomes a major new component of the modern society, as a platform for commerce, entertainment, etc., the potential impact may be very significant.

In our work, we are assisted by a large number of volunteers, including scientists, technologists, and educators, most of them professional members of MICA. Some of them are actively engaged in the VWs development activities under the auspices of various governmental agencies, e.g., NASA. We have also established a strong network of international partnerships, including colleagues and institutions in the Netherlands, Italy, Japan, China, and Canada (a list which is bound to grow). We are also establishing collaborative partnerships with several groups in the IT industry, most notably Microsoft Research, and IBM, and we expect that this set of collaborations will also grow in time. This broad spectrum of professionally engaged parties showcases the growing interest in the area of scientific and scholarly uses of VWs, and their further developments for such purposes.

**Acknowledgments.** The work of MICA has been supported in part by the U.S. National Science Foundation grants AST-0407448 and HCC-0917814, and by the Ajax Foundation. We also acknowledge numerous volunteers who have contributed their time and talents to this organization, especially S. McPhee, S. Smith, K. Prowl, C. Woodland, D. Enfield, S. Cianciulli, T. McConaghy, W. Scotti, J. Ames, and C. White, among many others. We also thank the conference organizers for their interest and support. SGD also acknowledges the creative atmosphere of the Aspen Center for Physics, where this paper was completed.